\documentclass[twocolumn,runningheads]{svjour2}
\smartqed  
\usepackage{graphicx}
\newcommand\sun{\odot}
\newcommand\arcdegr{\mbox{$^\circ$}} 
\newcommand\arcmin{\mbox{$^\prime$}}

\newcommand\aj{{AJ}}
\newcommand\apj{{ApJ}}
\newcommand\aap{{A\&A}}
\newcommand\mnras{{MNRAS}}

\journalname{Astrophysics and Space Science}
\begin{document}

\title{Gamma rays from colliding winds of massive stars
}


\author{Anita Reimer        \and
        Olaf Reimer         \and
        Martin Pohl 
}

\authorrunning{Reimer et al.} 

\institute{A. \& O. Reimer \at
              W.W. Hansen Experimental Physics Laboratory, Stanford University,
445 Via Palou, Stanford, CA 94305, USA \\
              \email{afr@stanford.edu, olr@stanford.edu}           
           \and
           M. Pohl \at
              Department of Physics and Astronomy,
   Iowa State University,
   Ames, Iowa 50011, USA \\
              \email{mkp@iastate.edu}  
}

\date{Received: date / Accepted: date}

\maketitle

\begin{abstract}
Colliding winds of massive binaries have long been considered as potential
sites of non-thermal high-energy photon production. 
This is motivated by the detection of non-thermal spectra in the radio band,
as well as by correlation studies of yet unidentified EGRET $\gamma$-ray sources with 
source populations appearing in star formation regions. 

This work re-considers the basic radiative processes and its properties that lead to high 
energy photon production in long-period massive star systems. We show that
Klein-Nishina effects as well as the anisotropic nature of the inverse Compton
scattering, the dominating leptonic emission process, likely yield spectral and variability 
signatures
 in the $\gamma$-ray domain at or above the sensitivity of
current or upcoming gamma ray instruments like GLAST-LAT.
In addition to all relevant radiative losses, we include propagation (such as
convection in the stellar wind)
as well as photon absorption effects, which a priori can not be neglected. 

The calculations are applied to WR~140 and WR~147, and predictions for 
their detectability in the $\gamma$-ray regime are provided. 
Physically similar specimen of their kind like 
WR~146, WR~137, WR~138, WR~112 and WR~125
may be regarded as candidate sources at GeV energies
for near-future $\gamma$-ray experiments.

Finally, we discuss several aspects relevant for 
eventually identifying this source class as a $\gamma$-ray emitting population.
Thereby we utilize our findings on the expected radiative behavior
of typical colliding wind binaries in the $\gamma$-ray regime as well
as its expected spatial distribution on the $\gamma$-ray sky. 

\end{abstract}

\section{Introduction}
\label{intro}

By far the most convincing evidence for particle acceleration to 
relativistic energies mediated by the supersonic (terminal velocity $v_{\infty}\sim$1000-5000~km/s) 
winds of massive ($\dot M\sim 10^{-6...-5}M_\sun$/yr),
hot ($T\sim 30000-50000$K) stars 
comes from the observation of non-thermal radio emission 
(e.g. \cite{Abbott86}).
This has been interpreted by synchrotron emission on the basis of the
measured spectra (much steeper than the canonical value $\alpha_r\sim +0.6$,
$F_\nu\propto \nu^{\alpha_r}$) and high 
brightness temperatures of $\sim 10^{6-7}$K, far exceeding $\sim 10^4$K
expected from free-free emission from a steady-state isothermal radially 
symmetric wind \cite{Wright1975}. Those particles have been suggested 
to be accelerated
either in shocks caused by the instability of radiatively driven winds
\cite{White1985}, in the shocked wind collision region of multiple systems
or in the termination shock \cite{Voelk1982}. Triggered by these observations it has been 
quickly realized that $\gamma$-ray emission should be expected as well, either
through leptonic processes (inverse Compton scattering (IC) of the copious
stellar UV photons (e.g. \cite{ChenWhite1991}), relativistic bremsstrahlung 
(e.g. \cite{Pollock1987}) or hadronic
interactions of co-accelerated ions with the dense wind material 
(e.g.\cite{White1992,Torres2004}).
This has established a plausible physical setting for massive star systems being
putative $\gamma$-ray emitters. Indeed, 
positional coincidences of Wolf-Rayet (WR) stars with the population of so far 
unidentified EGRET sources have been found for 13 WR-binary systems 
\cite{Romero1999,bene2001,bene2005}. 
Recently, the presence of non-thermal radio emission has been linked
to the binarity status of the stellar systems \cite{Dougherty2000},
which supports the picture of particles being predominantly accelerated
at the forward and reverse shocks from the colliding supersonic 
winds from massive stars.

In this work we consider long-period binary systems as the most prospective
$\gamma$-ray emitters detectable by the near-future experiment GLAST-LAT, and
 extend previous theoretical work by including propagation, and an\-iso\-tropy and 
Klein-Nishina (KN) effects of the inverse Compton (IC) scattering process. 

\section{The broadband SED of colliding wind systems of massive stars}

At present, the observationally established broadband spectral energy distribution (SED)
of colliding wind binaries (CWBs) ranges from radio wavelengths to X-ray energies. At GHz frequencies, only
one third of all observed WRs has been detected (e.g. \cite{Leitherer1995,Leitherer1997}). 
Roughly 40\% of these show signatures of a contributing non-thermal component.
These are predominantly binaries with periods $>1$~year, and this can be understood as 
an absorption effect: The winds of massive stars are partially optically thick
at radio wavelengths. For typical O-star wind parameters the $\tau_\nu=1$ surface
lies roughly at $\sim(1-2)\times 10^3$R$_\sun$ at GHz frequencies.
For WR stars these radii are even larger. Thus optically thin sight lines can only be 
found in systems with stellar separations of $\ll~10^{3-4}$R$_\sun$. At IR wavelengths,
massive binaries often show components in excess to the expected free-free emission
(e.g. \cite{Williams1990}).
These persistent or variable/periodically/episodical\-ly appearing IR excess fluxes have been 
interpreted as signatures of dust formation. The photospheric continuum emission
dominates a broad range at optical to UV energies. 
Excess emission has also been observed
at X-ray energies. Some systems show X-ray luminosities up to two orders of magnitude
above the values suggested by the canonical relation $L_X/L_{\rm bol}\sim 10^{-7}$
obeyed by single early-type stars (e.g. \cite{Garmany1991}). This excess emission
has been attributed to the shock-heated ($T\sim 10^{7-8}$K) plasma produced in the 
winds collision region. Indeed, recent Chandra observations of the close-by
WR 147 system show extended emission peaking at a location which is in agreement
with the expected wind collision region \cite{Pittard2002}.
The observed phase-locked orbital variability in massive binaries in the soft X-ray 
as well as radio band is mainly a result of absorption in the dense stellar winds 
along the changing line of sight.

In the following we shall concentrate on the non-thermal part of
the continuum emission from CWBs.

\section{The model}

Various works have been devoted in the past to non-thermal emission 
components from CWBs 
(e.g. \cite{ChenWhite1991,White1992,Usov1992,Eichler1993,bene2003}, see also references 
in \cite{review}) based on either leptonic (mostly IC) and/or hadronic 
($\pi^0$-decay $\gamma$-rays) processes. They generally conclude on $\gamma$-ray 
luminosities in 
the range $10^{32\ldots 35}$erg/s to be expected from WR-binaries.
Since the IC process has been shown to likely dominate the $\gamma$-ray production
(e.g. \cite{Eichler1993,Muecke2002}), this process is prone to determine the 
spectral appearance of WR-binaries at high energies. Here Klein-Nishina
as well as anisotropy effects, neglected in past modeling of these systems,
may provide valuable features that can be instrumental in identifying this
source population at $\gamma$-ray energies. Further spectral imprints are expected
from particle propagation within the extended colliding wind region. 
A simplified geometry of the system
proofs sufficient to highlight these effects.

In the following we consider
the sketch of a colliding wind region (CWR) that has been presented
by e.g. Eichler \& Usov (1993) \cite{Eichler1993} with the stagnation point defined by balancing 
the wind momenta
with the assumption of spherical homogenous winds. Since in general $\dot M_{WR}>\dot M_{OB}$
and $v_{\infty,WR}\approx v_{\infty,OB}$ the shock distance to the OB-star, $x_{OB}$, is
much smaller than to the WR-star.
We shall neglect here the interaction of the stellar radiation fields on the wind structure
(\cite{Gayley1997,Stevens1994}):
This effect is inherent to short-period binaries, and may influence the wind speed, thereby 
weakening the ram balance, shock strength and temperature. Thus our theoretical
considerations here shall be restricted to long-period binaries.
The stellar winds are permeated by magnetic 
fields originating from the surface of the massive star. Estimates for surface magnetic
field strengths range from below $B_s=100$~G (e.g. \cite{Mathys1999}) up to $\sim 10^4$G in WR-stars
\cite{Ignace1998}. In the following we fix this value to a reasonable 100~G, unless 
stated otherwise, and use the well-developped magnetic rotator theory (e.g. \cite{Weber1967}) to 
estimate
the field strength $B_G$ (in Gauss) at the CWR. Typically $>$mG or higher 
field strengths are expected, assumed to
be constant throughout the emission region, at the CWR in long-period binaries.
The shocked high-speed winds are creating a region of hot gas that is separated by a 
contact discontinuity, and a forward and reverse shock follows. The gas flow velocity in this 
region away from the stagnation point
will be some fraction of the wind velocity which we keep constant at $V$ for simplicity.
A simplification of the geometry
from a bow-shaped to a cylinder-shaped collision region (with radius $r$ perpendicular to the
line-of-centres of the two stars, and given thickness)
allows us to solve the relevant continuity equations analytically (see \cite{Reimer2006}).
Here we consider diffusive shock acceleration (with acceleration rate $a$) out of a 
pool of thermal particles, and
take into account (continuous) radiative losses (synchrotron, IC, bremsstrahlung and 
Coulomb losses), 
(energy-independent) diffusion by introducing an escape time $T_0$ and convection with 
speed $V$ (set to $V=1/2 v_{\rm OB}$ if not noted otherwise). The maximum particle energy is
thereby determined self-consistently.

\begin{figure}
\centering
  \includegraphics[height=6.5cm]{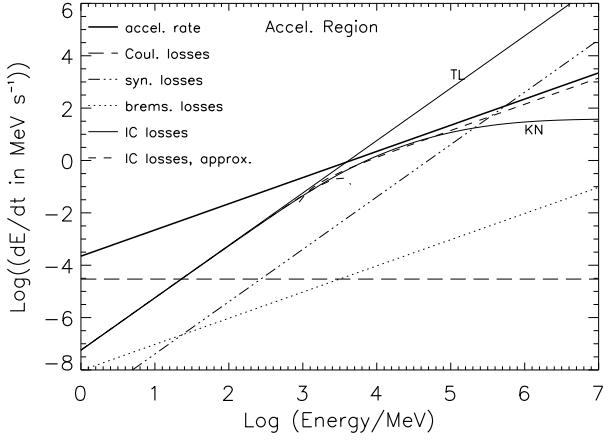}
\caption{Energy loss rates in the acceleration zone due to IC scattering (thin solid lines) in the Thomson regime (TL) and
KN regime (KN), synchrotron radiation (dashed-triple-dotted line), relativistic electron-ion
bremsstrahlung (dashed-dotted line) and Coulomb interactions (dashed line) in comparison 
to the acceleration rate
(thick solid line). Parameters are: bolometric OB-star luminosity $L_{\rm bol,OB}=10^5 L_{\sun}$,
target photon energy $\epsilon_T=10$~eV, OB-star mass loss rate $\dot M_{\rm OB}=10^{-6} M_{\sun}$yr$^{-1}$, 
WR-star mass loss rate $\dot M_{\rm WR}=10\dot M_{\rm OB}$,
OB-star terminal velocity $v_{\infty,\rm OB}=4000$km/s,
binary separation $D=10^{14}$cm, $x_{\rm OB}\approx 0.24D$, surface magnetic field $B_s = 100$G, 
field strength at the CWR $B\approx 0.5$G,
diffusion coefficient $\kappa_a = 1.6\cdot 10^{20}$cm$^2$s$^{-1}$, escape time $T_0\approx 1127$~s,
size of the acceleration region $r_0\approx 8.5\cdot 10^{11}$cm. See \cite{Reimer2006} for details.}
\label{fig:1}       
\end{figure}

At a distance $>r_0$ from the stagnation point convection along the post-shock flow
will dominate over diffusion. At $r_0$ diffusion balances convection.
Correspondingly, the emission region is devided into a region where acceleration/diffusion dominates,
the "acceleration zone", and the outer region where convection dominates, the "convection zone".
The steady-state diffusion-loss equation can be solved analytically provided suitable approximations
for the KN cross section are applied (see \cite{Reimer2006}). Fig.~\ref{fig:1} shows an example
of all loss time scales for typical parameters of WR-binaries. The transition region,
between the Thomson and extreme KN range of the IC cross section, appears relevant for
typical long-period WR-systems. A rigorous treatment of the Compton losses must therefore
include KN effects.

\begin{figure}
\centering
  \includegraphics[height=6.5cm]{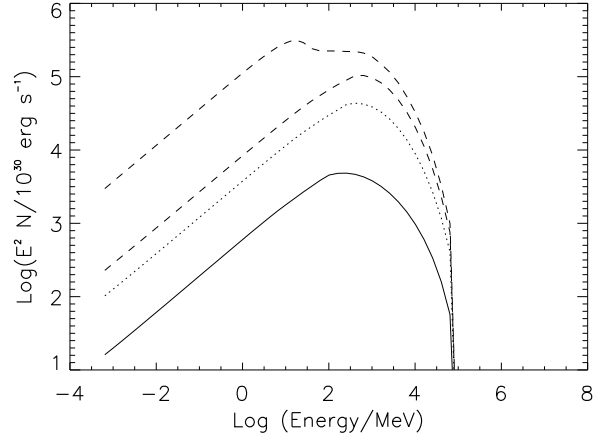}
\caption{IC spectra from the acceleration region for $D=10^{14}$cm and 
$\theta_{\rm L}=0\arcdegr$ (lower solid line), $30\arcdegr$ (lower dashed line),
$60\arcdegr$ (dotted line), $90\arcdegr$ (dashed-dotted line), $120\arcdegr$ (dashed-triple dotted line) $150\arcdegr$ 
(long dashed line)
and $180\arcdegr$ (upper solid line). 
For $\theta_{\rm L}=30\arcdegr$ the total volume-integrated (i.e. acceleration plus 
convection zone) IC spectrum is also shown (upper dashed line).
All other parameters are the same as in Fig.~\ref{fig:1}.}
\label{fig:2}       
\end{figure}

We find electron spectra with smooth roll-overs that cutoff at higher
energies as compared to electron spectra that remain in the Thomson approximation throughout
the whole particle energy range. 
In the convection region the particles loose energy radiatively as well as through expansion
losses while in the post-shock flow. This leads to a
dilution of the particle density as well as a deficit of high energy particles.
The corresponding volume-integrated photon spectra softens (see Fig.~\ref{fig:2}), 
depending on the relative 
size of the convection region with respect to the acceleration region. This itsself depends on the
diffusion and convection time scales operating in the CWR.

With typically the photon to magnetic field energy density 
$u_{\rm ph}/u_B\approx 52 L_{\rm bol,OB,\sun}/(B_G x_{\rm OB,\sun})^2\geq 1$ 
(where $L_{\rm bol,OB,\sun}$ and $x_{\rm OB,\sun}$ are
in units solar luminosity and radius, respectively)
IC scattering 
turns out to be the most important radiative loss channel for
relativistic electrons in CWBs in most cases \cite{Eichler1993,Muecke2002}.
Because the stellar target photons for IC scattering arrive at the collision region from a 
prefered direction, the full angular dependence of the cross section has to be taken into account 
(e.g. \cite{Reynolds1982}).
This leads to anisotropy effects like the emitted flux and cutoff energy 
being dependent on the sight line into the wind (see Fig.~\ref{fig:2}). The calculations are carried out
using the monochromatic approximation for the photospheric target field and an isotropic
particle distribution in the emission region. The latter is maintained by pitch angle scattering
at magnetic field inhomogeneities. Fig.~\ref{fig:2} shows the maximum flux level reached 
at phases where the WR-star is behind the OB-star along the sight line. Furthermore, 
large scattering angles tend to produce more energetic photons than low angles. 
This can affect the cutoff photon energy by several 
orders of magnitude for a given value of the electron energy cutoff (see Sect.~4.2). 

Details are described in \cite{Reimer2006}.

\section{Application to archetypal systems}

\subsection{WR 140}
\begin{figure}
\centering
  \includegraphics[height=6cm]{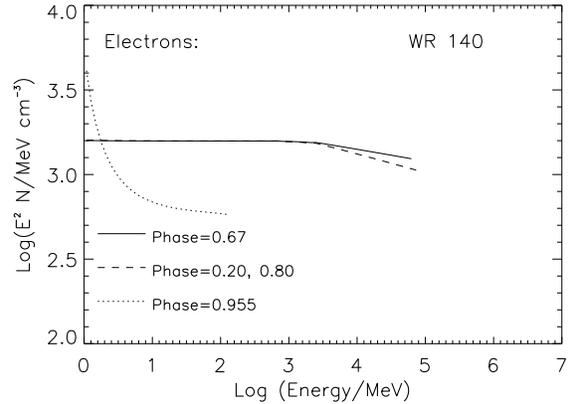}
\caption{Steady-state electron spectra for WR~140 at orbital phases 0.955, 0.2, 0.671 and 0.8. Flux
is in arbitrary units. See text for parameters.}
\label{fig:3}       
\end{figure}

\begin{figure}
\centering
  \includegraphics[height=6.5cm]{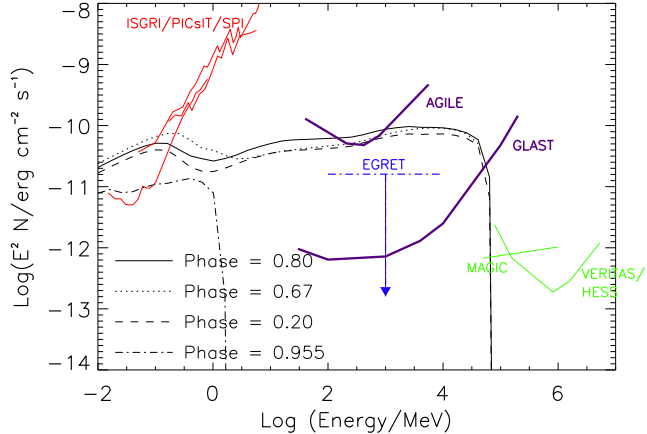}
\caption{
IC spectra for WR~140 at phases 0.955, 0.2, 0.671 and 0.8 from electron spectra as shown in
Fig.\ref{fig:3}. The spectral changes from $\gamma$-ray absorption are not shown here.
The EGRET $2\sigma$ upper limit \cite{Muecke2002} is based on observations that correspond to a superposition 
of orbital states rather determined by periastron phase, and therefore applies rather to phase 0.955.}
\label{fig:4}       
\end{figure}

The archetypical WR-binary system WR~140 (WC 7pd + O4-5 V), located in the Cygnus 
constellation at a distance of 
$\sim 1.85$ kpc, is one of the most detailled studied specimen of its kind. 
Its long ($\sim 8$~years) period and extreme eccentricity ($e\approx 0.88$) makes it 
a diverse system to study its non-thermal behavior in the radio band.
For the first time, Dougherty et al. (2005) \cite{Dougherty2005} resolved the bow-shaped arcs of the emission region at 
8 epochs of 8.4 GHz VLBA observations. 
We used the synchrotron spectra and system parameters
as published in \cite{Dougherty2005} to predict its non-thermal high energy emission at various
phases. With these values the CWR is located at a distance $x_{\rm OB}$ of 0.32 times the 
stellar separation.
For a surface magnetic field of 100 G roughly equipartition values for its field strength
at the CWR location follow. With a relativistic particle injection energy of $\sim 10^{-2...-3}\%$ 
of the kinetic wind energy,
the synchrotron fluxes at the considered phases (0.2, 0.67, 0.8, 0.95) could be reproduced.
From radio observational grounds, relativistic electrons at least up to $\sim$10-100 MeV do exist. 
This limits
the diffusion coefficient $\kappa$ to sufficiently low values which will allow the acceleration rate to
 overcome the Coulomb loss rate at low energies. 
A further constraint of the diffusion coefficient is provided by the spectral shape of
the non-thermal radio component, if the shock compression ratio and convection velocity
are known. On the other side, $\kappa$ is limited by Bohm diffusion.
For $\kappa=2\cdot 10^{19}$cm$^2$s$^{-1}$
up to $\sim 10^5$MeV-electrons are expected at least at apastron, where radiative losses are 
sufficiently small
not to affect the cutoff particle energy (see Fig.~\ref{fig:3}). 
The self-consistent determination of the maximum particle energy allows sound predictions 
at the highest energies.
Note that KN effects will alter the emitting electron spectrum above $\sim 10^4$MeV, as indicated
in Fig.~\ref{fig:3}. This is in contrast to phases close to periastron.
The intense stellar radiation field there limits the electron spectrum to $\sim 100$MeV 
(see Fig.~\ref{fig:3}). As a consequence, the corresponding photon spectrum from IC scattering
cuts off already
in the soft $\gamma$-ray band (see Fig.~\ref{fig:4}), and hadronically produced photons may
dominate at (sub-)GeV energies, albeit with a possibly undetectable flux level even for the
more sensitive near-future experiments.
EGRET observations, that were carried out rather 
around periastron, therefore did not lead to detections. 
Mainly as a result of including also hadronic ion-ion interactions into the calculations for
$\pi^0$ production, the corresponding $\pi^0$-decay $\gamma$-ray flux in Pittard \& Dougherty 
\cite{Pittard2006} extends to higher energies than considered here.
The association of WR~140 to the unidentified EGRET source 
3EG J2022+4317 appears to be vague: With WR~140 being located $\sim 0.67\arcdegr$
away from the nominal position of 3EG J2022+4317, and barely consistent with the 99\% source location
uncertainty contour a conclusive identification seems farfetched. 
It is due to the extreme eccentricity
 of this system, leading to significant changes in the stellar radiation field density
at the shock location, that causes a blurring of the phase-locked flux variations expected otherwise
from the anisotropy of the dominating IC scattering process. 
Indeed, orbital flux variations in this system are reduced
to a factor $\sim 2-3$ (see Fig.~\ref{fig:4}).
In general, relativistic bremsstrahlung radiation lies always 
below the IC emission level in these systems. When compared to the expected IC flux, 
$\pi^0$-decay $\gamma$-ray production is small, even if
the total wind energy is transformed into relativistic protons/ions.
This makes WR-binary systems rather unpromising putative neutrino and cosmic ray sources within our model.

Fig.~\ref{fig:4} implies that WR~140 may be detectable with GeV-instruments like GLAST-LAT even
at individually selected phases if the
electrons reach sufficient high energies, while INTEGRAL requires Msec exposures 
for any detection above 1~MeV. The expected extent of the photon spectrum to $\sim 10-100$~GeV
at phases where the binary separation is large, may even allow low-energy threshold Imaging Atmospheric
Cherenkov Telescopes (IACTs)
to gain important information on the cutoff of WR~140's spectrum. This may
allow to further constrain the diffusion coefficient. 
Note that photon absorption due to $\gamma\gamma$ pair production can not a priori be neglected
in colliding wind systems. E.g. for WR~140 the optical depth 
$\tau_{\gamma\gamma}(r=r_0,E\approx 100$GeV)$\sim 1$ at phase 0.67, 
$\tau_{\gamma\gamma}(r=r_0,E\approx 100$GeV)$\sim 3$ at phase 0.95.

\subsection{WR 147}

\begin{figure}
\centering
  \includegraphics[height=6.5cm]{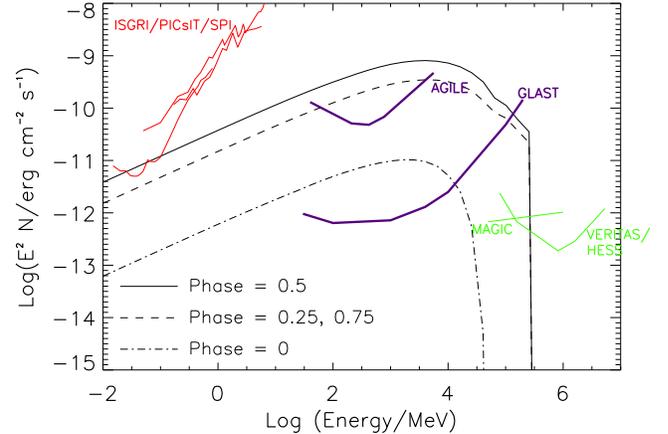}
\caption{
IC spectra of WR~147 for orbital phases 0, 0.25, 0.5 and 0.75. 
See text for parameters. $\gamma\gamma$ pair production 
absorbs not more than $\leq$0.3\% ($>50$~GeV) and $\leq$18\% ($>100$~GeV) of the produced flux at
orbital phases 0.25 and 0.5, respectively (not shown in figure). No absorption takes place
at phase 0.}
\label{fig:5}       
\end{figure}

Due to its proximity WR~147 (WN8h + B0.5V) is one of the few binary systems where high resolution
radio observations lead to resolving this system into its components (Williams et al. 1997).
The observed radio morphology supports a binary separation of $\sim 417$~AU for a
source distance of 650~pc. We used the system parameters as published in
Setia Gunawan et al. (2001) \cite{Setia2001} together with the observed synchrotron spectrum to model
its expected spectral behaviour at high energies. Since WR~147's inclination $i$ nor
eccentricity $e$ are known, we assumed for our modeling $e=0$ and $i=90\arcdegr$.
The observed synchrotron spectrum could be reproduced within its observational
uncertainties if $\sim 0.15\%$ of the OB-wind kinetic energy is transformed into
relativistic electrons, and assuming a surface magnetic field of 30~G (translating into 25~mG fields
at the CWR). Due to its huge binary separation the emitting particle spectra are not limited
by IC losses but rather by the size of the acceleration region. Fig.~\ref{fig:5} presents the expected
orbital IC flux variations due to the anisotropic nature of the IC process for the
above described parameter set. Here the maximum flux and photon cutoff energy, expected when the 
WR-star is behind the OB-star
along the sight line, are more than one order of magntitude higher than their minimum values.
Absorption of $>50$~GeV photons turns out negligible for this system,
provided its eccentricity is small.
The chance to detect WR~147 with future high sensitive instruments is somewhat better than for
WR~140. While LAT will probe the MeV-GeV range of its spectrum,
only the low threshold ($<100$ GeV) experiments in the 
northern hemisphere among the ground-based IACTs may have a distinct chance to 
detect WR~147 at its highest energies.

\subsection{Further candidates ?}
 
Our work provides ample evidence
for the archetypal long-period WR-binaries WR~140 and WR~147 being
plausible candidates for detection with near future $\gamma$-ray instruments.
Both provide target photon fields at the CWR which is
sufficiently dense to produce a IC scattered photon flux above the respective instrument sensitivity,
but at the same time does not act as a thick absorber for those photons, nor
prevents electron acceleration to high enough energies by radiative losses.
A sensible search for further $\gamma$-ray source candidates may therefore
start with scanning the WR-binary population for physically similar systems.
A (not complete) list encompasses WR~146, WR~137, WR~112 and WR~125, with
WR~112 being the most distant ($\sim 4$~kpc) of them.
From all of them non-thermal radio components have been observed
at least occasionally.
One (WR~137) has been found positionally coincident with an
unidentified EGRET source (3EG J2016+3657: \cite{Romero1999}).

\section{Perspectives for the gamma-ray band}

The identification of a specific colliding wind system or the population thereof as 
$\gamma$-ray emitters requires first: a convincing positional association, 
and second: its physical relation with the detected $\gamma$-ray source.

The most up-to-date catalog of WR-stars in the Milky Way lists more than
220 sources with a detected binary frequency of $\sim 40-50\%$ in the solar 
neighborhood \cite{WRcat}. 
Its spatial distribution reflects the spiral arm structure of our Galaxy with a slight asymmetry
with respect to the galactic plane (\cite{WRcat} and references therein).
64 WR-(systems) are found in the solar neighbourhood ($<2$~kpc), more than 25 have been detected
in the immediate vicinity ($<30$pc) of the galactic center. The total number of
WR-(systems) in the Galaxy has been estimated to up to $\sim 8000$ \cite{Prantzos1986},
total OB-star number counts in our Galaxy may reach values as high as 60000 \cite{Reed2000}.
This is orders of magnitude larger than the number of all, presumably galactic, but
still unidentified $\gamma$-ray sources detected to date.

Intriguing positional coincidences of massive star populations, among them colliding wind systems,
with unidentified EGRET sources have been
found in the past (e.g. \cite{Romero1999,bene2001}), however no individual
binary system could unambigeously be identified as a high energy $\gamma$-ray emitter so far. 
This is most likely a statistical issue. EGRET's large location error box, typically
$\sim 0.25\arcdegr$ for a strong point source at high latitudes, and even larger for sources
in the galactic plane, allows to find many plausible counterparts within the positional uncertainty
of the $\gamma$-ray source. Furthermore, the strong diffuse $\gamma$-ray background at low galactic
latitudes makes the detection of any non-periodic galactic source close to the instrument sensitivity
even more challenging.
The capabilities of EGRET's successor, GLAST-LAT
\footnote{http://www-glast.slac.stanford.edu/software/IS/glast\_lat\_performance.htm}, 
will improve here. 
Its expected ability of both detecting weak signals ($>3\cdot 10^{-9}$cm$^{-2}$ s$^{-1}$) and 
at the same time
locating an object in the $\gamma$-ray sky within $\sim 0.5\arcdegr-5\arcmin$, solves the
confusion problem at least among the galactic bright EGRET sources.

How will a typical CWS appear at $\gamma$-ray energies? What are its observable characteristics
that may discriminate it from the wealth of other plausible $\gamma$-ray source classes?

To begin with, the available energy kinetic wind energy $L_w=\dot M v_\infty^2/2$ provides a firm upper 
limit on the photon production rate from each process, and thus on the expected flux level.
Inserting typical mass loss and wind parameters one finds the kinetic energy of massive star winds
to be of order 1\% of the bolometric radiative energy output, or typically $\sim 10^{38}$erg/s. 
Taking into account the efficiency of the particle acceleration process, much smaller values
will rather be the rule.

Furthermore, colliding wind regions will appear not extended,
but point-like on the $\gamma$-ray sky for both present and near future satellite and 
ground based instruments. 

Our study implies that detectable phase-locked orbital variations 
in the $\gamma$-ray band occur, however, they may possibly be blurred by the geometry of the system
(in particular due to eccentricity and inclination). Flux variations at high energies
may additionally be expected from modulations of the geometry and size of the emitting region,
modulations of the stellar target photon field density, inhomogeneities
and/or shocks in the wind outflow, etc. The time scale of these variations is determined
by the respective physical cause.

Spectral hints on the expected {\it low-energy} IC component of CWBs may be deduced from
their visible synchrotron spectrum.
Spectral indices at GHz frequencies in the range $\alpha_r$=0.3...-1.3 has been 
detected so far from the current 
limited sample of non-thermal emitters among the WR-binary systems. This may hint towards
photon spectra with photon index $\alpha_{\rm ph}<2.5$ far below the cutoff energy.
At GeV-energies KN, anisotropy and propagation effects may have an impact on the spectral shape.

With these characteristics at hand, a population study may be instrumental
to finally unveil the class of massive star binary systems as high energy emitters.

\begin{acknowledgements}
We like to thank the organizers of the conference ``The multi-messenger approach to 
high energy gamma-ray sources'', where this article was presented, for a very 
fruitfull meeting. 
\end{acknowledgements}



\end{document}